\documentstyle[12pt]{article}

\font\tenrm=cmr10

\newcommand{\bref}[1]{(\ref{#1})}
\newcommand{\ct}[1]{\cite{#1}}

\newcommand{\be}{\begin{equation}}
\newcommand{\ee}{\end{equation}}

\newcommand{\beq}{\begin{equation}}
\newcommand{\eeq}{\end{equation}}

\def\lsim{\mathrel{\rlap{\lower4pt\hbox{\hskip1pt$\sim$}}
    \raise1pt\hbox{$<$}}}
\def\gsim{\mathrel{\rlap{\lower4pt\hbox{\hskip1pt$\sim$}}
    \raise1pt\hbox{$>$}}}
\def\frac#1#2{{{#1} \over{#2}}}

\begin{document}
\begin{titlepage}

\begin{flushright}
{LBNL-52402 \\ }
{\hfill March 2003 \\ }
\end{flushright}
\vglue 0.2cm

\begin{center}
{
{On the Speed of Gravity and 
the $v/c$ Corrections to the Shapiro Time Delay \\ }
\vglue 1.0cm
{Stuart Samuel$^{1}$ \\ }
\vglue 0.5cm

\vglue 0.4cm
{\it Theory Group, MS 50A-5101 \\}
{\it Lawrence Berkeley National Laboratory\\}
{\it One Cyclotron Road\\}
{\it Berkeley, CA 94720 USA\\}

\vglue 0.8cm


{\bf Abstract}
}
\end{center}
{\rightskip=3pc\leftskip=3pc
\quad Using a relatively simple method, 
I compute the $v/c$ correction to the gravitational time 
delay for light passing by a massive object moving with speed $v$. 
It turns out that the $v/c$ effects are too small 
to have been measured in the recent 
experiment involving Jupiter and quasar J0842+1845 
that was used to measure the speed of gravity. 

\medskip
\medskip

\noindent
{\it PACS indices:} 04.20.Cv, 04.80.Cc, 04.25.Nx, 98.54.Aj, 96.30.Kf
}

\vfill          

\textwidth 6.5truein
\hrule width 5.cm
\vskip 0.3truecm
{\tenrm{
\noindent
$^{1)}$\hspace*{0.15cm}E-mail address: samuel@thsrv.lbl.gov \\ 
}}

\eject
\end{titlepage}

\newpage

\baselineskip=20pt

{\bf\large\noindent I.\ Introduction}\vglue 0.2cm

On September 8, 2002, 
a conjunction of quasar J0842+1835 and Jupiter took place. 
This event was used to measure the Shapiro time delay of 
the quasar signal due to the gravity of 
Jupiter.\ct{fomalontkopeikin,fomalontkopeikin2} 

Many years ago, 
I.\,I.\,Shapiro proposed one of the classic tests 
of general relativity in which radio signals are bounced off 
an inner planet during a superior conjunction 
with the Sun.\ct{shapiro}
The effect of the Sun's gravity is to create a delay in the time 
required for the radio waves to return to Earth. 
In subsequent years, 
measurements performed using Mercury 
confirmed Einstein's theory, 
and  
the PPN parameter $\gamma$ was measured to be its expected 
value of $1$ to within 10\%.\ct{shapiroetal68,shapiroetal71} 

Because Jupiter's gravity is weaker than the Sun's, 
the QSO J0842+1835 measurement required remarkable accuracy: 
$10^{-12}$ seconds. 
This was achieved using very long baseline interferometry.
Motivation for undertaking this experiment 
stems from a proposal\ct{kopeikina}
that it can be used 
to measure the speed of gravity $c_g$. 
The idea of testing whether $c_g$ equals the speed of light $c$, 
as should be the case in general relativity, 
has attracted considerable attention 
both 
in the astrophysics community\ct{asada,kopeikinb,will,kopeikinc,faber} 
and in the media\ct{nyt}. 
The measurement  yielded  
$c_g/c = 1.06 \pm 0.21$\ct{fomalontkopeikin,fomalontkopeikin2} 
and was hailed as 
a confirmation of Einstein's general theory of relativity.

The purpose of this {\it{Letter}} is to point out an error 
in the theoretical formula used to analyze 
the Jupiter/quasar experiment 
and to provide the correct result. 
In reference\ct{kopeikina,kopeikinb}, 
a $v/c_g$ correction 
to the Shapiro time delay in the Jupiter/quasar experiment
is found to be proportional to $1/\theta^2$, 
where $\theta$ is the angle between the 
quasar and Jupiter.  
Since $\theta$ is small, 
an enhancement occurs 
thereby making the measurement feasible. 
However, 
using a simple method, 
this {\it Letter} 
computes the $v/c$ corrections 
and finds no such term. 
The discrepancy between the formula 
of the current work and
the one used in the experiment is understood: 
The angle $\theta$ in the latter
was actually not the observable one 
but an artificially defined angle. 

Our notation conforms to that of 
references \ct{kopeikina} and \ct{kopeikinb}, 
which are henceforth indicated as A and B: 
Quasar J0842+1835 is located in the direction 
of the unit vector $\vec K$. 
See Figure 1. 
Radiation for the quasar arrives at two observational 
points $1$ and $2$ on Earth, 
which are separated from one another by the distance $\vec B$. 
The impact parameters 
for each of these two points is respectively denoted by 
$\vec \xi_1 = \xi_1 \vec n$ and $\vec \xi_2 = \xi_2 \vec n$. 
Here, $\vec n$ is a unit vector perpendicular to $\vec K$ 
going from Jupiter to the closest approach 
of the electromagnetic radiation of the quasar. 
Since the difference of the impact parameters is small 
compared to either impact parameter,
we use $\xi$ to denote the value of either  
when a distinction is not important. 
The velocity of Jupiter is indicated as $\vec v_J$, 
and  
the Earth-Jupiter distance is denoted by $R_{EJ}$. 

We are interested in the most significant corrections 
to the Shapiro time delay for the Jupiter/quasar experiment. 
Therefore, we neglect terms proportional to the product of two of, 
or the square  
of, any of the following small, dimensionless quantities: 
$ { {G_N M_J } \over {\xi c^2} } \approx 6 \times 10^{-9}$, 
${ {v_J} \over {c} } \approx 4.5 \times 10^{-5}$, 
${ {B} \over {\xi} } \le 0.006$ and 
$\theta_{obs} = { {\xi} \over { R_{EJ} } } \sim 0.001 $ 
(which is the angle that an astronomer observes between 
Jupiter and the quasar). 
Here, 
$G_N$ is Newton's constant and $M_J$ is the mass of Jupiter. 

\medskip

{\bf\large\noindent II.\ The $v_J/c$ Corrections}\vglue 0.2cm

If $\Delta t_1$ and $\Delta t_2$ denote 
the Shapiro time delays at the points $1$ and $2$, 
then the quantity of interest is the difference 
$\Delta \left( t_1 , t_2 \right) = \Delta t_2 - \Delta t_1$:
\be
 t_2 - t_1 = 
    | \vec x_2 (t_2) - \vec x_0 |/ c - 
    | \vec x_1 (t_1) - \vec x_0 |/ c + 
      \Delta \left( { t_1 , t_2 } \right)
\quad . 
\label{0}
\ee 
Here, $t_1$ and $t_2$ are respectively the times at which the signals 
are measured at the two points $\vec x_1 (t_1)$ and $\vec x_2 (t_2)$ 
on Earth, 
$\vec x_0$ is the position of the quasar,  
and 
$ | \vec x_2 (t_2) - \vec x_0 |/ c - | \vec x_1 (t_1) - \vec x_0 |/ c$ 
is the time difference that occurs when gravitational effects 
are absent. 

If $\vec B = \vec x_2 (t_2) - \vec x_1 (t_1)$ 
and $\vec n$ are oppositely oriented, 
or more precisely $\vec B \cdot \vec n < 0$,
then $\xi_1 > \xi_2$ and $\Delta \left( t_1 , t_2 \right)$ 
is positive because the electromagnetic radiation that arrives 
at $2$ undergoes more time delay because it passes closer 
to Jupiter.
This is the case illustrated in Figure 1. 

If Jupiter were not moving, 
which is the static situation, 
then the Shapiro time delay for a single wave is\ct{weinberg} 
\be
  \Delta t = {{2G_N M_J} \over {c^3}}\left( {1+\ln \left( {{{4R_{JQ} R_{EJ}} \over {\xi^2}}} \right)} \right) 
\quad ,  
\label{0p5} 
\ee
where $R_{JQ}$ is the distance from Jupiter to the quasar. 
The leading contribution to $\Delta \left( t_1 , t_2 \right)$ is 
therefore 
\be
  \Delta \left( {t_1,t_2} \right) = 
  \Delta t_2 - \Delta t_1 = 
  {{4G_N M_J} \over {c^3}}\ln \left( {{{\xi_1} \over {\xi_2}}} \right) 
    ={{4G_N M_J\Delta \xi } \over {\xi c^3}} 
\quad .  
\label{1}
\ee

Let us determine $\Delta \xi = \xi_1 - \xi_2$ in terms of $\vec B$. 
The electromagnetic rays that originate from the quasar 
are bent slightly as they pass by Jupiter by 
an amount $\Delta \varphi$ given by\ct{weinberg} 
$$
  \Delta \varphi = {{4G_N M_J} \over {\xi c^2}}
\quad . 
$$ 
Eventually, one finds
\be
  -\vec n\cdot \vec B = 
  \Delta \xi \left( {1+{{4G_N M_J R_{EJ}} \over {\xi^2 c^2}}} \right) 
  \approx \Delta \xi = \xi_1 - \xi_2 
\quad ,  
\label{2}
\ee 
because 
$$
  {{4G_NM_JR_{EJ}} \over {\xi^2 c^2}} 
   \le {{4G_NM_JR_{EJ}} \over {R_J^2c^2}} \sim 0.001 
\quad ,  
$$ 
where $R_J$ is the radius of Jupiter. 
In other words, 
within the solar system 
the angular deflection created by Jupiter can be neglected,  
and the separation between the rays remains essentially constant. 

By substituting Eq.\bref{2} into \bref{1},  
one obtains the result for a static Jupiter 
\be
  \Delta \left( {t_1,t_2} \right) 
  = -{{4G_N M_J\vec n\cdot \vec B} \over {\xi c^3}} 
\quad .  
\label{3}
\ee 
When $\xi = \theta_{obs} R_{EJ}$ is used 
in Eq.\bref{3}, 
it reproduces the leading term in the notation of references A and B. 

Let us now compute the $v_J/c$ corrections. 
This is simple to do by selecting 
an appropriate reference frame. 

During the time 
in which the rays propagate from Jupiter to the Earth, 
Jupiter moves almost in a straight line with constant speed. 
In other words, 
the orbital motion of Jupiter around the Sun is not important. 
The same is true for the Earth. 
Therefore, observers on both planets can be considered 
as being inertial. 
Let us select an observational frame 
for which Jupiter is motionless. 
In this frame, 
the Earth appears to be moving with a velocity $\vec v_E$ 
equal to $- \vec v_J$. 
Since Jupiter is not moving, 
Eq.\bref{3} applies. 
However, the distance $\vec B_{sf}$ between points 1 and 2 
as measured in the static frame 
is not equal to $\vec B$ as measured on Earth. 
Place a static observer at the point $1$ at time $t_1$ 
and another static observer at the point $2$ at time $t_2$. 
Have these observers make the time measurements. 
Then the situation is completely static and 
the formulas for the static case may be used. 

During the time $t_2 - t_1$,  
the Earth moves a distance $\vec v_E \left( t_2  - t_1 \right)$.
Next, 
note that the leading contributions to 
$t_2  - t_1$ are 
\be
  t_2 - t_1 \approx 
   - { { \vec K \cdot \vec B } \over {c} } 
   + { { \vec n \cdot \vec B \theta_{obs} } \over {c} }  
   + \Delta \left( t_1 , t_2 \right),
\quad ,   
\label{3p5}
\ee 
of which the first is the largest. 
Therefore, 
\be
  \vec B_{sf} = \vec B  
   - { { \vec K \cdot \vec B } \over {c} } \vec v_E 
     + { { \vec n \cdot \vec B \theta_{obs} } \over {c} } \vec v_E
     + \Delta \left( {t_1,t_2} \right) \vec v_E
\quad .  
\label{4}
\ee 

The motion of Earth leads to two corrections to the static 
time delay difference in Eq.\bref{3}. 
Using $\vec n\cdot \vec B_{sf}$ 
in Eq.\bref{3} leads to an additional term 
$ { { 4G_N M_J \vec n \cdot \vec v_E \vec K \cdot \vec B } 
    \over {\xi c^4} }$. 
The other correction arises 
if the Earth moves toward (or away from) Jupiter. 
In this case, 
the time delay is reduced (or increased) by the time 
$\delta \Delta \left( {t_1,t_2} \right)$ 
it takes light to travel the distance determined by the difference 
between $\vec B_{sf}$ and $\vec B$. 
The corresponding correction due to the second and third terms 
of Eq.\bref{4}
is independent of $G_N$ and is a contribution to 
the first part of Eq.\bref{0} 
that involves the difference in distances between 
the positions of the quasar and 
the observation points $1$ and $2$. 
The fourth term in Eq.\bref{4}
leads to  
$$
  \delta \Delta \left( {t_1,t_2} \right) = 
    - { {\vec K \cdot \vec v_E} \over {c} } \Delta \left( {t_1,t_2} \right) 
$$
One switches to the Earth frame using $\vec v_E = - \vec v_J$. 
The final result is 
\be
  \Delta \left( {t_1,t_2} \right) = 
    -{ {4 G_N M_J}  \over {\xi c^3} } 
   \left( { 
      \vec n \cdot \vec B
      \left( { 1 + { {\vec K \cdot \vec v_J} \over {c} } } \right) + 
      { {\vec K \cdot \vec B \vec n \cdot \vec v_J} \over {c} } 
   } \right)
\quad .  
\label{5}
\ee

The correction factor ${{{\vec K \cdot \vec v_J} \over c}}$ 
is present in references A and B. 
However, 
we find no $1/\theta^2$ terms. 
In its place is the 
${\vec K \cdot \vec B \vec n \cdot \vec v_J} / {c}$ term 
of Eq.\bref{5}. 

Although the Shapiro time delay 
has effects 
created by the long-ranged gravitational force 
(e.g.\,see Eq.\bref{0p5}), 
these effects cancel in the time difference 
of Eq.\bref{0}. 
In the static case, 
this is illustrated by Eq.\bref{3}, 
in which $\Delta \left( { t_1, t_2 } \right)$ 
is expressed in terms of the impact parameters 
of the electromagnetic waves, 
that is, quantities measurable in the vicinity of Jupiter. 
One therefore expects that long-ranged effects 
should not be present in $\Delta \left( { t_1, t_2 } \right)$ 
even in the non-static case. 
The $1/\theta_{obs}^2$ terms of reference A and B, 
however, grow with the Earth-Jupiter distance. 
On physical grounds, 
it seems unlikely that such terms are present, 
and our computation confirms this. 

The leading term in $\Delta \left( {t_1,t_2} \right)$ 
is of the order of $100$ nanoseconds for the Jupiter/quasar experiment.
The $v_J/c$ corrections 
in Eq.\bref{5}   
are at least 10,000 times smaller making them less than 
$0.01$ nanoseconds. 
It is therefore impossible that 
references \ct{fomalontkopeikin} and \ct{fomalontkopeikin2} 
measured the speed of gravity. 
The $v_J/c$ corrections are also masked by larger corrections 
such as terms down by $B/\xi$ and $\theta_{obs}$, 
which are present but not shown in this {\it Letter}. 

\medskip 

{\bf\large\noindent III.\ Comparison to References A and B}\vglue 0.2cm

It is easy to find the source of the $1/\theta^2$ effects 
in references A and B. 
In those works, 
the times 
$s_1 = t_1 - | \vec x_1 \left( {t_1} \right) - \vec x_J \left( {s_1} \right)|/c$ 
and 
$s_2 = t_2 - | \vec x_2 \left( {t_2} \right) - \vec x_J \left( {s_2} \right)|/c$ 
at which rays $1$ and $2$ 
pass by Jupiter are expanded in terms of the times 
$t_1$ and $t_2$ when the rays are observed at the points 
$\vec x_1$ and $\vec x_2$ on Earth. 
The differences between the $s_i$ and $t_i$ are sizeable, 
of order of $R_{EJ}/c$,  
and during this time, 
Jupiter moves a significant distance. 
See the dotted circle in Figure 1. 
References A and B 
define the angle $\theta$ 
in terms of the position of Jupiter at $t_1$. 
This is not the physically observed angle $\theta_{obs}$. 
For clarity, 
denote the angle of references A and B by $\theta_{AB}$. 

When the electromagnetic waves from the quasar pass by Jupiter, 
sunlight that has been reflected off of Jupiter also heads toward Earth. 
Eventually, 
the various waves arrive on Earth. 
See Figure 1. 
It is evident 
that the angle $\theta_{obs}$ 
between the quasar and Jupiter 
observed by an astronomer on Earth 
is determined by Jupiter's position 
at time $s_1$ and not $t_1$. 

The reason for the  $1/\theta^2$ term 
in references A and B 
is due to the use of the artificial angle $\theta_{AB}$. 
The relation between $\theta_{obs}$ and $\theta_{AB}$ is
\be
   \theta_{obs} \approx \theta_{AB} + { {\vec n \cdot \vec v_J } \over {c} }
\quad .  
\label{6}
\ee
When this result is 
substituted into the leading term of Eq.\bref{3}, 
\be
  \Delta \left( {t_1,t_2} \right) 
  = -{{4G_N M_J \vec n \cdot \vec B} \over { R_{EJ} \theta_{obs} c^3}} 
  = -{{4G_N M_J \vec n \cdot \vec B} \over { R_{EJ}  c^3}} 
    \left( { 
   {{1} \over {\theta_{AB}} } - 
   { {\vec n \cdot \vec v_J } \over {c \theta_{AB}^2 } }  
           } \right)
\quad ,   
\label{7}
\ee
and the $1/\theta^2$ effect emerges. 

References A and B express $\Delta \left( {t_1,t_2} \right)$ as 
\be
  \Delta \left( {t_1,t_2} \right) = 
  {{4G_N M_J } \over { c^3}} \ln { 
  \left[ {
      { r_{1J}\left( {s_1} \right) + 
      \vec K\cdot \vec r_{1J}\left( {s_1} \right) }
     \over {r_{2J}\left( {s_2} \right) + 
      \vec K\cdot \vec r_{2J}\left( {s_2} \right)} 
          } \right] }
\quad ,   
\label{8}
\ee
and then expands unwisely about $t_1$. 
The expansion is somewhat subtle since factors such as 
$ 
   r_{1J} \left( {s_1} \right) + 
      \vec K\cdot \vec r_{1J}\left( {s_1} \right) 
$
are proportional to the small quantity $\theta_1^2$. 
A careful analysis reveals that Eqs.(8) and (29) of 
references A and B should have used  
\be
  \ln { 
  \left[  {
      { { r_{1J}\left( {s_1} \right) + 
      \vec K\cdot \vec r_{1J}\left( {s_1} \right) }
     \over 
        { r_{2J}\left( {s_2} \right) + 
      \vec K\cdot \vec r_{2J}\left( {s_2} \right) } 
      }    } \right] } 
  =   \left[ {
      { { r_{1J}\left( {t_1} \right) + 
      \vec K\cdot \vec r_{1J} \left( {t_1} \right) }
     \over 
        { r_{2J}\left( {t_1} \right) + 
      \vec K\cdot \vec r_{2J} \left( {t_1} \right) } 
       }
             } \right]  + 
  { { 2 \vec n \cdot \vec v_J \vec n \cdot \vec B } \over 
     { c r_{1J} \theta_{AB}^2 } }
\quad .   
\label{9}
\ee
When Eq.\bref{9} is substituted into Eq.\bref{8}, 
the $1/ \theta_{AB}^2$ term of Eq.\bref{7} due to the 
expansion in Eq.\bref{6} is reproduced. 
This shows that our equations are consistent with the method 
used in references A and B. 

Summarizing, 
(1) an analysis using a static frame allows 
one to easily compute the $v_J/c$ corrections 
from the static result 
and one finds no $1/\theta_{obs}^2$ terms, 
(2) physical considerations suggest 
that such terms are absent, 
(3) an ill-advised 
expansion about the arrival time $t_1$ in references A and B  
produces the 
artificial $1/\theta_{AB}^2$ effects, 
and 
finally, (4) consistency with references A and B is achieved 
after the expansion in Eq.\bref{9} is used. 

There is nothing wrong with using Eq.\bref{7} 
as long as the instantaneous angle $\theta_{AB}$ 
of Figure 1 is used. 
However, in the analysis of data,
reference \ct{fomalontkopeikin2} 
begins with Eq.\bref{7} 
and replaces $c$ with $c_g$. 
This is a mistake since the difference between the angles 
$\theta_{obs}$ and $\theta_{AB}$ is due to the motion 
of Jupiter during the period in which light travels 
from Jupiter to the Earth 
and has nothing to do with gravity. 
It is not surprising therefore that when a fit 
to Eq.\bref{7} is performed 
with $c_g$ replacing $c$ that the 
experimentally deduced value of $c_g$ is approximately $c$. 
Reference \ct{fomalontkopeikin2} is only measuring 
the leading contribution to the Shapiro time delay 
in rearranged form as must be the case since the $v_J/c$ 
effects were beyond detection with current radio telescopes.  

\medskip 

{\bf\large\noindent IV.\ On the Notion of the Speed of Gravity}\vglue 0.2cm

It is clear from the derivation using the static frame 
in Section II that 
the leading $v_J/c$ corrections involve the speed 
of light and not the speed of gravity, 
and there is a recent analysis\ct{will}  
that supports this claim. 
See also references \ct{asada} and \ct{faber}. 
However, 
references A and B argue that 
$v_J/c_g$ should appear. 
The issue here is how does one extend Einstein's general theory 
of relativity to allow the possibility that 
the speed of gravity $c_g$ is not equal to $c$. 
A reasonable approach is to assume that the effect of gravity 
propagates at $c_g$ instead of $c$. 
For example, 
in the retarded times and positions of Jupiter in formulas, 
one replaces $v_J/c$ by $v_J/c_g$.
Hence, 
in the frame in which Jupiter is moving and the Earth 
is at rest, 
the $v_J/c$ effects are generated 
in the vicinity of Jupiter, 
and
$v_J/c_g$ should appear in lieu of $v_J/c$
in Eq.\bref{5} of
$\Delta \left( { t_1 , t_2 } \right)$. 
But consistancy demands 
that the computation of 
$\Delta \left( { t_1 , t_2 } \right)$ 
be frame independent. 
Thus, 
there does not seem to be a consistant way to define 
the speed of gravity concept for the Jupiter/quasar experiment. 
In the static frame, 
the corrections are due to the speed of light, 
while in the Jupiter-moving frame they are due to the speed of gravity. 

How then might one try to test $c_g \ne c$ 
in Einstein's theory of relativity?
The static and Jupiter-moving frames 
are both inertial. 
If Jupiter happened to be accelerating toward (or away from) 
the quasar's electromagnetic waves as they passed by the planet, 
then one would not be able to go back and forth 
between the two frames. 
Therefore, there is a reasonable chance that 
the speed of gravity concept could be defined for such a situation. 
The parameter $c_g$ would not be attached to velocity-dependent terms 
but to acceleration effects. 
Although it is worth exploring this possibility theoretically, 
it is unlikely that a system within or beyond our Solar System 
exists that generates an effect sufficiently large to be 
measurable with current instruments.

\medskip

{\bf\large\noindent Acknowledgments}

This work was supported in part
by the Director, Office of Science, 
Office of High Energy and Nuclear Physics, 
of the Department of Energy 
under contract number DE-AC03-76SF00098.


\medskip 
{\bf\large\noindent Figure Captions}  

\noindent
Figure 1. The Motion of Electromagnetic Waves Relevant 
for the Jupiter/Quasar Experiment. \\
For clarity, the diagram is not drawn to scale.

\vfill\eject
\end{document}